\documentclass[aps,twocolumn,showpacs,prl]{revtex4}
\usepackage{graphicx}
\usepackage{bm}
\usepackage{amsmath}
\usepackage{amsfonts}
\newcommand{\ch}{{\cal H}}

\newcommand{\tr}{{\rm Tr}}
\newcommand{\ket}[1]{| #1 \rangle}
\newcommand{\bra}[1]{\langle #1 |}
\newcommand{\braket}[2]{\langle #1 | #2 \rangle}

\begin{document}
\title{Security of entanglement-based quantum key distribution with
practical detectors}

\author{Masato Koashi, Yoritoshi Adachi, Takashi Yamamoto, and Nobuyuki Imoto}
\affiliation{Division of Materials Physics,
Department of Materials Engineering Science,
Graduate School
of Engineering Science, Osaka University, 
1-3 Machikaneyama,
Toyonaka, Osaka 560-8531, Japan}
\affiliation{CREST Photonic Quantum Information Project, 4-1-8 Honmachi, Kawaguchi, Saitama 331-0012, Japan}

\begin{abstract}
We prove the unconditional security of an entanglement-based
 quantum-key-distribution protocol using detectors that respond to
 multiple modes of light and cannot distinguish 
between one from two or more photons. 
Even with such practical
 detectors, any defect in the source is automatically detected as an
increase in the error rate or in the rate of double clicks.
\end{abstract}
\maketitle

The idea of using quantum entanglement for 
absolutely secure secret communication was first proposed 
by \mbox{Ekert} \cite{ekert91:_quant_theor}, followed by 
a proposal of a modified quantum-key-distribution (QKD)
protocol (BBM92)
by Bennett {\it et al.} 
\cite{bennett92:_quant_theor}.
When ideal apparatuses are used and the source is 
possessed by a legitimate user, the BBM92 protocol
is equivalent to the BB84
protocol \cite{Bennett-Brassard84}, which does not use an 
entangled source. On one hand, this property has lead to 
a powerful security proof \cite{Shor-Preskill00} 
based on entanglement, which is 
applicable to prepare-measure protocols such as
the BB84 protocol \cite{Shor-Preskill00}
and the B92 protocol \cite{Bennett92,TKI03,Koashi04}.
But on the other, the equivalence
 may have discouraged the use of an entangled source in 
an actual setup if the same function is available without
the trouble of generating entanglement.
In fact, a huge advantage of actually using 
an entangled source shows up when we take 
defects in the source into account. Defects 
may arise from limitation on technology, and 
in the BB84 protocol
they raise new threats on the security such as 
the photon-number-splitting attack \cite{BLMS00}.
Even worse, in long-distance communication a
source must be placed at an insecure relay station
and hence its property cannot be trusted anymore.
The entanglement-based protocol such as the BBM92 protocol
provides a unique property in this situation.
Since the protocol is based on testing 
a strong correlation unique to the entanglement,
we may expect that any defect in the source will 
be revealed as a degradation of the correlation.

An important question at this point is what kind of 
detection apparatus is required to realize such a 
built-in mechanism for detecting the defects in the 
source. It would surely be a disappointment if we were 
forced to use an ideal detector for such a purpose.
So far, it has been shown \cite{Koashi-Preskill03} that 
it is sufficient if
one of the two parties have a detection apparatus with 
a so-called squashing property \cite{GLLP02}, that is,
equivalence to a noisy quantum channel followed by an 
ideal BB84 measurement on a qubit. It is expected that 
a practical detection apparatus (as in Fig. 1) with two 
threshold (on/off) detectors will satisfy the squashing 
property if we assign a random bit whenever both detectors
 have clicked. Based on this conjecture, practical benefits 
of the BBM92 protocol, such as placing the source in the 
middle to achieve a larger communication distance,
were discussed \cite{MFL07} quantitatively. 
But the proof of the conjecture remains open, leaving 
an unsatisfactory situation that the BBM92 protocol,
being one of the basic QKD
protocols with many experimental demonstrations 
\cite{exp1,PFUBLMPSKWJZ04,UTSWSLBJPTOFMRSBWZ07,exp2},
still requires an assumption in the detectors for its security.

In this paper, we prove the unconditional security of 
the BBM92 protocol with practical threshold detectors 
which cannot distinguish between one photon from two or more, 
and cannot single out a single optical spatio-temporal mode either.
Instead of proving the squashing property, we adopt a protocol 
in which the double-click events are simply discarded.
The proof is based on a simple
 inner-product formula for the basis states, which shows that 
the parity of the number of incident photons has an important role.
Eve can carry 
out a powerful attack by distributing
 odd and even numbers of photons to the two receivers.
The security is essentially obtained by monitoring the bit-error rate 
and the double-click rate to watch out for the possibility of such an
attack.

\begin{figure}
\center{\includegraphics[width=.95\linewidth]{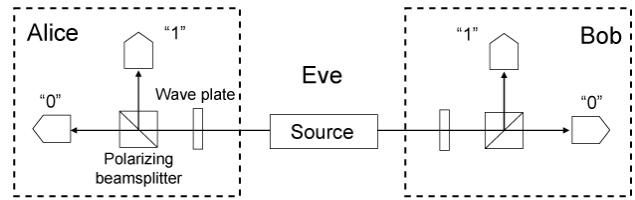}}
\caption{Schematic of the setup for the BBM92 protocol.  
\label{fig:setup}
}
\end{figure} 

The protocol considered here is the BBM92 protocol with the detection
apparatuses shown in Fig.~1. For each event, Alice randomly chooses 
between the $Z$ basis and the $X$ basis using a wave plate placed before 
the polarizing beam splitter (PBS). In the $Z$-basis measurement,
horizontal ($H$) and vertical ($V$) polarization components 
are split at the PBS and sent toward 
two threshold detectors corresponding to bit values 
0 and 1. In the $X$-basis measurement,
the $\pm 45^\circ$ polarizations ($D\pm$) are split instead.
Alice publicly announces whether she detected photons, and if so,
she also announces whether both of the detectors clicked (double clicks).
Bob follows the same protocol as Alice. 
The bit values are registered only when both parties have detected photons, 
but neither party has seen double clicks. 

As usual, we assume that non-unit efficiency and dark counting of the
detectors can be equivalently described by a noise source in front of
the detection apparatus. This is satisfied, for example, if two
detectors with matched efficiency are used and their roles are switched 
randomly. Hence, here and henceforth, we treat each detector as an ideal 
threshold detector that clicks when it receives one or more photons.

Let $N$ be the number of events where
both Alice and Bob detected photons and their basis choices
were the same. In principle, the number of photons 
($n_A\ge 1$) incident on Alice's apparatus can be determined for 
each event, since this observable commutes with Alice's actual 
measurement. The same goes with Bob's photon number $n_B\ge 1$.
Accordingly, the $N$ events are classified into  
$N\xi$ `multi-photon events' satisfying $n_A+n_B\ge 3$
and $N(1-\xi)$ `single-photon events' with $n_A=n_B=1$.
Among the $N\xi$ multi-photon events, suppose that
$N\xi\delta_m$ events were discarded due to double clicks, 
and $N\xi\epsilon_m$ events showed bit errors, namely, 
different bit values were registered by Alice and Bob.
The single-photon events should have no double clicks,
and suppose that they include $N(1-\xi)\epsilon_1$ 
bit-error events.
Whereas the parameters $(\xi,\delta_m,\epsilon_m,\epsilon_1)$ 
are all measurable in principle, 
the actual setup does not reveal $(n_A, n_B)$
and hence only tells us 
the overall double-click fraction $\delta$ 
and a good estimate of the overall error fraction $\epsilon$, which are 
related to $(\xi,\delta_m,\epsilon_m,\epsilon_1)$ as 
\begin{eqnarray}
 \delta &=& \xi\delta_m,
\label{eq:delta}
\\
 \epsilon &=& (1-\xi)\epsilon_1+\xi\epsilon_m.
\label{eq:epsilon}
\end{eqnarray}
 From the $N$ events, Alice and Bob produce sifted key of length 
$N(1-\delta)$ with a quantum bit error rate (QBER) $\epsilon/(1-\delta)$.
For simplicity,
we assume that the error correction is done by encrypted 
one-way communication 
from Alice to Bob by consuming the previously shared secret key 
of length 
$N(1-\delta)fH(\epsilon/(1-\delta))$, where 
$H(x)\equiv -x\log_2 x -(1-x)\log_2 (1-x)$ and 
$f\ge 1$ represents the 
inefficiency in the practical error correction schemes.
The reconciled key is further shortened by $N\tau$ to
amplify the privacy, where $\tau$
is determined from the observed values $(\delta,\epsilon)$.
The fraction $R_{\rm key}$ of the final key 
(normalized by $N$)
is thus written as follows,
\begin{eqnarray}
 R_{\rm key}=(1-\delta)\left[
1-fH(\epsilon/(1-\delta))\right]-\tau(\delta,\epsilon).
\label{eq:key-rate}
\end{eqnarray}
In the limit of large $N$, the final key is secure if 
\begin{eqnarray}
 \tau(\delta,\epsilon)\ge 
(1-\xi)H(\epsilon_1)+\xi(1-\delta_m)
\label{eq:tau-condition}
\end{eqnarray}
holds for any attack by Eve, because the right-hand side
is given by the argument by Gottesman {\it et al.} \cite{GLLP02} 
with a pessimistic assumption that Eve perfectly knows Alice's 
bit value in multi-photon events.
One might expect that the use of multi-photons inevitably 
leads to bit errors $\epsilon_m>0$
and double clicks $\delta_m>0$, but 
it turns out that either value can be zero by choosing a
suitable state. But Eve cannot make both of the values to be 
zero at the same time. In what follows, we determine this 
trade-off relation and determine  $\tau(\delta,\epsilon)$
satisfying Eq.~(\ref{eq:tau-condition}).

Let us first suppose that Alice (or Bob) receives $n$ photons
in a single spatio-temporal mode.
Let $\ket{Q,n}$ be the state with $n$ photons in the same polarization
$Q$, namely, $\ket{Q,n}\equiv (n!)^{-1/2}(a_Q^\dagger)^n\ket{vac}$
with $a_Q$ being the photon annihilation operator for polarization $Q$.
On the $Z$-basis, the outcome $0$ corresponds to 
the projection to the state 
$\ket{0_{Z}^{(n)}}=\ket{H,n}$,
and $1$ to the state $\ket{1_{Z}^{(n)}}=\ket{V,n}$.
The other $n-1$ orthogonal states correspond to the double clicks.
On the $X$-basis, the outcomes $0$ and $1$ correspond to
the states $\ket{0_{X}^{(n)}}=\ket{D+,n}$ and
$\ket{1_{X}^{(n)}}=\ket{D-,n}$. 
Using $a_{D\pm}=2^{-1/2}(a_H\pm a_V)$, 
 we obtain a relation vital to our discussion,
\begin{eqnarray}
 \braket{b_{X}^{(n)}}{b_{Z}^{\prime(n)}}=(-1)^{bb'n} 2^{-n/2}.
\label{eq:fundamental}
\end{eqnarray}
We can show that this relation is unaltered even if $n$ photons 
are distributed over multiple modes. In such a case, the photon
numbers $n_1,n_2,\ldots$ in each mode can be measured in principle. 
For fixed values of $\{n_j\}$, the state $\ket{0_{Z}^{(n)}}$
is given by $\ket{H,n_1}\ket{H,n_2}\cdots$, and so are the other three 
states. Noting that $\sum_j n_j=n$, one can see that the inner products
are still given by Eq.~(\ref{eq:fundamental}). The only difference is
the dimension $d=\prod (n_j+1)$ of the state space, but it does not affect
the argument below, in which only Eq.~(\ref{eq:fundamental}) is used.

When $n=2l+1 (l=1,2,\ldots)$, Eq.~(\ref{eq:fundamental}) reads
$\braket{b_{X}^{(2l+1)}}{b_{Z}^{\prime(2l+1)}}= 
\braket{b_{X}^{(1)}}{b_{Z}^{\prime(1)}} 2^{-l}$, which 
leads to a clear physical interpretation. Since the dimension 
$d$ is even, the state space $\ch_A$ can be identified with a combined 
system $\ch_{A'}\otimes \ch_{A''}$
of a single photon (qubit) $A'$ 
and an ancilla $A''$, with the relations
\begin{eqnarray}
&& \ket{b_{W}^{(2l+1)}}_A= \ket{b_{W}^{(1)}}_{A'}
\ket{\phi^{(l)}_W}_{A''} \; (W=Z,X),
\label{eq:odd-decomp}
\\
&& {}_{A''}\braket{\phi^{(l)}_X}{\phi^{(l)}_Z}_{A''}= 2^{-l},
\label{eq:odd}
 \end{eqnarray}
which preserve the inner products (\ref{eq:fundamental}).
Hence for an odd number of incident photons,
Alice's measurement can be regarded as an ideal BB84 
measurement on a qubit $A'$, except that the outcome is overridden 
by the occurrence of double clicks that is determined 
by a basis-dependent measurement on the ancilla $A''$.
On the other hand, for even numbers we have 
a {\it constant} inner product
\begin{eqnarray}
 \braket{b_{X}^{(2l)}}{b_{Z}^{\prime(2l)}}= 2^{-l},
\label{eq:even}
\end{eqnarray}
which has no simple connection to a qubit.

Now let us derive a trade-off relation between 
$(\delta_m, \epsilon_m)$. 
For the moment, we consider the attacks
using only a single combination of $(n_A,n_B)$. 
For each event, the measurement operator 
$F_{\rm err}$ for having a bit error 
and 
$F_{\rm cor}$ for sharing the same bit value
are given by
\begin{eqnarray}
 F_{\rm err}&=&2^{-1}\sum_{W=X,Z}\sum_{b=0,1}
  P(\ket{b_W^{(n_A)}}_A\ket{(1-b)_W^{(n_B)}}_B),
\label{eq:F-err}
\\
F_{\rm cor}&=&2^{-1}\sum_{W=X,Z}\sum_{b=0,1}
  P(\ket{b_W^{(n_A)}}_A\ket{b_W^{(n_B)}}_B),
\label{eq:F-cor}
\end{eqnarray}
where $P(\ket{\cdot})\equiv \ket{\cdot}\bra{\cdot}$.
$1-F_{\rm cor}-F_{\rm err}$ corresponds to double clicks.
Let us write an expectation value of observable $O$
as $\langle O \rangle_\rho\equiv \tr(O\rho)$.
If $r(\langle F_{\rm cor}\rangle_\rho,
\langle F_{\rm err}\rangle_\rho)\le 0$
holds for any state $\rho$, the probability 
of $(\delta_m, \epsilon_m)$ to be deviated from
the region $r(1-\delta_m-\epsilon_m, \epsilon_m)\le 0$
is exponentially small for large $N$
\cite{TKI03,BTBLR05}.
In what follows, we consider the limit $N\to \infty$ and 
ignore such rare possibilities.
We divide the argument according to the parities of $(n_A,n_B)$.

i) {\it Odd--odd}, $n_A=2l_A+1$ and $n_B=2l_B+1$ with $l_A+l_B\ge 1$.
From Eqs.~(\ref{eq:odd-decomp}), (\ref{eq:F-err})
and (\ref{eq:F-cor})
with $P_W\equiv P(\ket{\phi^{(l_A)}_W}_{A''}\ket{\phi^{(l_B)}_W}_{B''})$,
we have
\begin{eqnarray}
 F_{\rm cor}+F_{\rm err}
=1_{A'}\otimes 1_{B'}\otimes (P_Z+P_X)/2.
\end{eqnarray}
Eq.~(\ref{eq:odd}) shows that  
the largest eigenvalue of $P_Z+P_X$ is $1+2^{-l_A-l_B}$, and 
we have 
\begin{eqnarray}
 \delta_m \ge (1-2^{-l_A-l_B})/2\ge 1/4.
\label{eq:odd-condition}
\end{eqnarray}

ii) {\it Odd-even}, $n_A=2l_A+1\ge 1$ and $n_B=2l_B\ge 2$. 
According to Eq.~(\ref{eq:fundamental}), 
there exists a unitary $V$ satisfying 
\begin{eqnarray}
 V\ket{a_{W}^{(n_A)}}_A\ket{b_{W}^{(n_B)}}_B
= \ket{a_{W}^{(n_A)}}_A\ket{(b+a \;{\rm mod}\; 2)_{W}^{(n_B)}}_B
\label{eq:V}
\end{eqnarray}
for $W=X,Z$. The operation of $V$ is 
regarded as a basis-independent controlled-NOT gate,
which is possible because the target system $B$ 
is not a qubit but has a larger dimension.
Since Eve is allowed to prepare any state, 
it makes no difference if we assume that she applies
$V$ just before she sends systems $A$ and $B$ to Alice and Bob.
Then the relevant observables take simple forms as follows:
\begin{eqnarray}
 V^\dagger F_{\rm err}V&=&1_{A'}\otimes (P^1_Z+P^1_X)/2,
\label{eq:oddeven-Ferr}
\\
 V^\dagger F_{\rm cor}V&=&1_{A'}\otimes (P^0_Z+P^0_X)/2,
\label{eq:oddeven-Fcor}
\end{eqnarray}
where $P^b_W \equiv P(\ket{\phi_W^{(l_A)}}_{A''}\ket{b_{W}^{(2l_B)}}_B)$.
This leads to 
\begin{eqnarray}
 \epsilon_m \ge g(\delta_m) 
\;\;{\rm for} \;\; \delta_m \le 1/3,
\label{eq:even-condition}
\end{eqnarray}
where $g(\delta)\equiv [(1-\delta)/2] -\sqrt{\delta(1-2\delta)}$.
The boundary is achievable with $n_A=1$ and $n_B=2$.
Of course, the case with $n_A=2l_A\ge 2$ and $n_B=2l_B+1\ge 1$
follows the same condition. 

Incidentally, the existence of the operation $V$
leads to an interesting attack by Eve with $n_A=1$ and $n_B=2$. 
Suppose that Eve prepares a maximally entangled state 
$\ket{\phi^+}_{AE}$ and a pure state $\ket{\chi}_B$,
and then applies unitary $V$ before she distributes the photons 
to Alice and Bob. As is seen from Eqs.~(\ref{eq:oddeven-Ferr})
and (\ref{eq:oddeven-Fcor}), $(\delta_m, \epsilon_m)$ is 
determined solely by the state $\ket{\chi}_B$, and hence 
Eve can realize any point on the boundary $\epsilon_m = g(\delta_m)$
by choosing $\ket{\chi}_B$ to be 
$\sum_W \alpha \ket{0_W^{(2)}} + \beta \ket{1_W^{(2)}}$.
On the other hand, Eq.~(\ref{eq:V}) shows that
Alice's outcome can be regarded as obtained from the direct 
measurement on $\ket{\phi^+}_{AE}$. Hence after the basis is
announced, Eve precisely learns Alice's bit. 
This particular attack constitutes a lower bound 
$\tau_{\rm low}$ on 
$\tau(\delta,\epsilon)$ to have a secure key:
\begin{eqnarray}
\tau_{\rm low}(\delta,\epsilon) \equiv 
\max_{\xi} 
\left[
\xi -\delta
+ (1-\xi) H\left(
\frac{\epsilon-\xi g (\delta/\xi)}{1-\xi}
\right)
\right].
\end{eqnarray}

iii) {\it Even--even}, $n_A=2l_A\ge 2$ and $n_B=2l_B\ge 2$.
Let $\ket{\phi_W^{\pm}}_{AB}\equiv 2^{-1/2}[
\ket{0_{W}^{(n_A)}}_A\ket{0_{W}^{(n_B)}}_B
\pm \ket{1_{W}^{(n_A)}}_A\ket{1_{W}^{(n_B)}}_B]$
and define the projection
$P^{\phi\pm}_W\equiv \ket{\phi_W^{\pm}}\bra{\phi_W^{\pm}}$. 
Further let 
$\ket{\psi_W^{\pm}}_{AB}\equiv 2^{-1/2}[
\ket{0_{W}^{(n_A)}}_A\ket{1_{W}^{(n_B)}}_B
\pm \ket{1_{W}^{(n_A)}}_A\ket{0_{W}^{(n_B)}}_B]$
and define $P^{\psi\pm}_W$ accordingly. We see from Eq.~(\ref{eq:even})
that the state with the minus sign 
(such as $\ket{\psi_X^{-}}_{AB}$)
is orthogonal to 
 any of the other seven states. 
Hence we can write 
\begin{eqnarray}
 F_{\rm err}&=& 2^{-1}[(P^{\psi+}_Z+P^{\psi+}_X)
\oplus (P^{\psi-}_Z + P^{\psi-}_X) \oplus 0],
\\
 F_{\rm cor}&=& 2^{-1}[(P^{\phi+}_Z+P^{\phi+}_X)
\oplus 0 
\oplus (P^{\phi-}_Z + P^{\phi-}_X)].
\end{eqnarray}
This leads to the same condition as Eq.~(\ref{eq:even-condition}).

\begin{figure}
\center{\includegraphics[width=.68\linewidth]{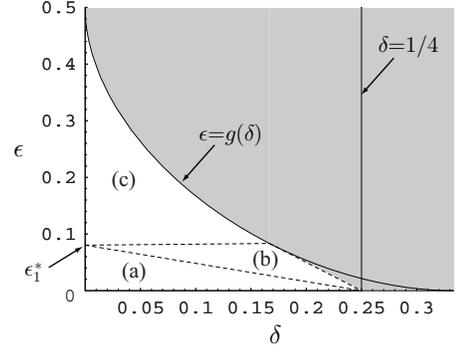}}
\caption{The observed fractions $(\delta, \epsilon)$
of double clicks and of bit errors 
 are a mixture of 
the multi-photon contribution (the shaded region) and 
the single photon contribution ($\delta=0$).
\label{fig:dbl-err}
}
\end{figure} 

Since the general attack is a mixture of attacks to various 
$(n_A,n_B)$, we conclude that $(\delta_m,\epsilon_m)$ 
must be in the shaded region of Fig.~2, obtained by taking 
convex combination of Eqs.~(\ref{eq:odd-condition}) 
and (\ref{eq:even-condition}).
$\tau(\delta,\epsilon)$ is then determined as the maximum of the 
right-hand side of Eq.~(\ref{eq:tau-condition})
 under the constraints Eqs.~(\ref{eq:delta}) and (\ref{eq:epsilon}).
The optimization is 
reduced to a standard problem of determining the convex hull of 
the points 
$(0,\epsilon_1,H(\epsilon_1))$ $(0\le\epsilon_1\le 1/2)$, 
$(\delta_m,g(\delta_m),1-\delta_m)$ $(0\le \delta_m \le 1/3)$,
and $(1/4,0,3/4)$. We classify the results into the three different 
regions (a)--(c) shown in Fig.~2.
Let $\epsilon_1^*\cong 0.080$ be the root of 
$16\epsilon_1^*(1-\epsilon_1^*)^3=1$. 

\noindent
(a) For $\epsilon\le \epsilon_1^*(1-4\delta)$,
\begin{eqnarray}
 \tau(\delta,\epsilon)=3\delta+(1-4\delta)H(\epsilon/(1-4\delta)).
\end{eqnarray}

\noindent
(b) For $\epsilon_1^*(1-4\delta)\le \epsilon \le \min\{
(1-6\delta)\epsilon_1^*+(\delta/2), 1/4-\delta\}$,
\begin{eqnarray}
 \tau(\delta,\epsilon)=[c_1\delta+c_2\epsilon+c_3]/(1-4\epsilon_1^*)
\end{eqnarray}
with constants $c_1\equiv 3-4H(\epsilon_1^*)+4\epsilon_1^*$, 
$c_2\equiv 4(1-H(\epsilon_1^*))$, and 
$c_3\equiv H(\epsilon_1^*)-4\epsilon_1^*$.

\noindent
(c) For $(1-6\delta)\epsilon_1^*+(\delta/2)\le \epsilon \le g(\delta)$,
\begin{eqnarray}
 \tau(\delta,\epsilon)=\tau_{\rm low}(\delta,\epsilon).
\end{eqnarray}

\begin{figure}
\center{\includegraphics[width=.83\linewidth]{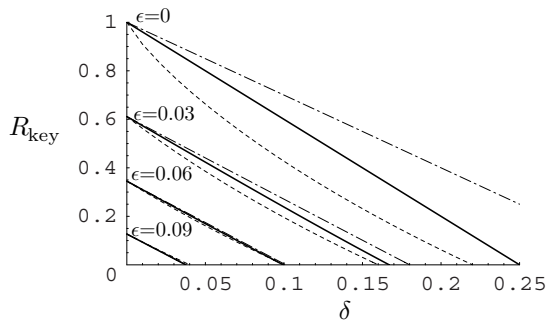}}
\caption{Dependence of key fraction $R_{\rm key}$ 
on the double-click fraction $\delta$. Solid curves are 
proved to be secure in the discarding protocol. Dash-dotted 
curves are the upper bounds for the discarding protocol.
Broken curves are the key fraction conjectured to be secure in the 
random-bit-assignment protocol. 
\label{fig:key-rates}
}
\end{figure} 

Figure 3 (solid curves) shows the key fraction $R_{\rm key}$
in Eq.~(\ref{eq:key-rate})
assuming the ideal error correction ($f=1$).
We see that the key fraction has almost linear dependence 
on the double-click fraction $\delta$.
The dash-dotted curves are the key fractions assuming 
$\tau(\delta,\epsilon)=\tau_{\rm low}(\delta,\epsilon)$,
which is the upper bound on the key fraction for any protocol
in which the privacy amplification for the single-photon events 
costs $H(\epsilon_1)$. The difference is not so large, indicating 
that the pessimistic condition (\ref{eq:tau-condition}) 
we used for simplifying argument does not sacrifice the efficiency much.

For comparison, we have added to Fig.~3 the broken curves 
$R_{\rm key}=1-2H(\epsilon + \delta/2)$, which is the key fraction 
conjectured to be secure in the protocol with random-bit assignment
for the double-click events.
For lower values of $\epsilon$, we see that 
discarding the double-click events 
is better than assigning a random bit and raising the error rate
as a result. When $\epsilon$ is larger, both curves are almost the
same. Hence for almost all practical purposes, the random-bit assignment
is unnecessary. On the other hand, from the theoretical point of view, 
it is interesting to notice that 
the conjectured curve for high $\epsilon$ slightly 
exceeds even the upper bound on the discarding protocol.
This may suggest that keeping Eve uninformed about
the occurrence of double clicks could have an advantage even at the 
cost of raising the error rate by the random-bit assignment.
For definite answers, we must wait for the development of the 
security analysis for the random-bit-assignment protocol
\cite{footnote}.

To conclude, we have proved the unconditional security of 
an entanglement-sharing QKD protocol (the BBM92 protocol)
with the use of practical detection apparatuses
and with no assumption on the source, which establishes  
the prominent feature of the protocol --- 
the built-in mechanism for detecting defects in the source.
We chose to discard 
the double-click events, which enabled us to build up the proof 
from a very simple nonorthogonality relation 
[Eq.~(\ref{eq:fundamental})] that holds regardless of the mode structure
of incident photons.
The proved secure key rate is higher than or almost the same as 
the rate conjectured for the random-bit-assignment protocol,
and hence practical benefits of the BBM92 protocol discussed by 
Ma {\it et al.} \cite{MFL07} are now confirmed with unconditional security.
The security proof is also applicable to a long-distance QKD 
using quantum repeaters \cite{BDCZ98}.

The authors thank K. Azuma for helpful 
discussions. 
This work was supported by MEXT Grant-in-Aid 
for Young Scientists (B) 17740265.


\begin{thebibliography}{22}
\bibitem{footnote}
At the time of writing, an interesting 
approach \cite{Tsurumaru-Tamaki08} came to our notice, in which 
a weaker version of the squashing property is proved, and 
is claimed to be enough to prove the security
for single-mode inputs.
\end{thebibliography}

\end{document}